\documentclass[10pt]{article}
\usepackage[margin=0.82in]{geometry}
\usepackage{amsmath,amssymb,amsthm,mathtools,mathrsfs}
\usepackage{float}
\usepackage{bm}
\usepackage{braket}
\usepackage{physics}
\usepackage{booktabs}
\usepackage{array}
\usepackage{enumitem}
\usepackage{graphicx}
\usepackage{todonotes}
\usepackage[
    colorlinks=true,
    linkcolor=red,
    citecolor=red,
    urlcolor=red
]{hyperref}
\usepackage[nameinlink,noabbrev]{cleveref}
\usepackage{tcolorbox}

\title{\textbf{Private communication from Pauli channels with no privacy}}
\author{
  Uthirakalyani G\thanks{These authors contributed equally.},
  Pritam Halder\footnotemark[1],
  David Elkouss\thanks{Corresponding author:
    \texttt{david.elkouss@oist.jp}.}\\[0.5em]
  \small Networked Quantum Devices Unit, Okinawa Institute of Science and Technology Graduate University, Okinawa, Japan}
\date{}

\begin{document}
\maketitle

\begin{abstract}
A channel capacity quantifies the communication capability of a noisy physical process. In contrast to communication channels in the classical world, quantum theory makes this capability contextual. We show that two Pauli two-qubit channels, each with zero private classical capacity, can be used to transmit private information when used together. One is a two-qubit Pauli channel whose environment can reconstruct the receiver output up to matrix transposition; the other is an antidegradable channel. We obtain a similar result when the second channel is the 50\% qubit erasure channel. A simple binary code built from rank-three mixtures of Bell states activates private communication. The main ingredient in our construction is a transpose-antidegradable channel that is not antidegradable.
\end{abstract}

\section{Introduction}
The capacity of a classical channel for transmitting information is given by a simple optimization formula over a single use of the channel \cite{shannon1948mathematical}. Its solution determines the usefulness of the channel and zero capacity certifies a complete lack thereof. A similar statement holds if we are interested in sending private information to a distant party over a wiretap channel \cite{csiszar1978broadcast}. Here we show that two quantum channels with zero capacity for private classical communications can transmit private communication when coupled together; in striking contrast with their classical counterparts. 

The analogous phenomenon for transmitting quantum information, called superactivation of quantum capacity \cite{Smithyard}, combines channels whose individual capacities vanish for different reasons. One channel is a private Horodecki channel $\mathcal{N}_H$ \cite{horodecki2005secure}; that is a channel that distributes states that are too noisy to be distilled but nonetheless can be used to send private information: $P(\mathcal N_H)>Q(\mathcal N_H)=0$ . The other is an antidegradable channel, a channel where the environment can simulate the output of the channel. That is given a channel $\mathcal N $, there exists a map $\mathcal{D}$, acting on the complementary channel $\mathcal{N}^c$ such that $\mathcal{D}\circ\mathcal N^c=\mathcal N$. Zero capacity follows from a simple no-cloning argument. The joint use of the two zero-capacity channels can nevertheless transmit quantum information. The main ingredient of the proof is the relation
\begin{equation}
\label{eq:smithyard}
    \frac{1}{2}P(\mathcal N)\leq Q(\mathcal N\otimes\mathcal A)
\end{equation}
which holds for arbitrary $\mathcal N$ and where $\mathcal A$ is a symmetric channel of unbounded dimension.

Whether a similar behavior holds for private classical communication was explicitly left open \cite{Smithyard}. The first obstacle is finding suitable channels to combine. One possible choice is an antidegradable channel, as they also have zero private capacity. Short of a pair of zero-private-capacity channels, Li, Winter, Zou, and Guo proved that private capacity is nonadditive using a zero-private-capacity erasure channel~\cite{li2009private}, while Smith and Smolin exhibited extensive nonadditivity, with an arbitrarily large gain from combining a channel of small private capacity with a zero-private-capacity channel~\cite{smith2009extensive}. 

Antidegradable channels can be generalized as follows: Suppose the environment is at least as informative as the receiver about every classical input label, including when both outputs are compared with the same arbitrary quantum reference correlated with the input. This property is called anti-complete less-noisiness (anti cLN) and it is implied by antidegradability. It induces a partial ordering between channels which is preserved under tensor products, so pairs of channels satisfying it cannot superactivate one another \cite{hirche2022contraction}.  Superactivation consequently requires at least one zero-private-capacity channel outside this class \cite{hirche2022bounding}. In particular, this argument discards leveraging the recent milestone construction of a non-antidegradable channel with zero private capacity \cite{chengkai2026quantum} since it is anti cLN.

Our construction uses two ingredients: an antidegradable channel and a channel for which the environment can reconstruct the output \textit{up to matrix transposition}, also known as a transpose-antidegradable channel~\cite{siddhu16}.  
We follow the terminology of Singh and Datta~\cite{Singh_2022}, distinguishing the linear-transpose notion from the original conjugate formulation~\cite{bradler10}.
\section{A transpose antidegradable Pauli channel}
Let $\mathcal N$ be a quantum channel from a sender to a receiver. Its complementary channel, denoted by \(\mathcal N^c\), describes the output delivered to the environment. The channel $\mathcal N$ is transpose-antidegradable if there exist a completely positive, trace-preserving map \(\mathcal D\) such that

\begin{equation}
\mathcal D\circ\mathcal N^c=\mathsf T\circ\mathcal N,
\end{equation}
where \(\mathsf T\) is matrix transposition in a fixed output basis. 
Consider the following channel on two qubits:

\begin{equation}
\mathcal F(\rho)
=\frac16\sum_{P\in\mathscr S}P\rho P^{\dagger},
\qquad
\mathscr S=\{X\otimes X,\,Y\otimes I,\,I\otimes Z,
\,X\otimes Z,\,I\otimes Y,\,Y\otimes Z\}.
\end{equation}

Here \(X,Y,Z\) are the Pauli operators and \(I=I_2\) is the single-qubit identity. An interesting property of this channel is that pairwise products of distinct elements of $\mathscr S$ generate all nonidentity two-qubit Paulis: given a two-qubit Pauli $P\neq I_4$, there is a unique unordered pair $Q,R\in\mathscr S$ such that $P=QR$ up to a phase (see Table \ref{tab:six-pauli-products}).
\begin{table}[ht]
\centering
\renewcommand{\arraystretch}{1.2}
\begin{tabular}{c|cccccc}
  & $XX$ & $YI$ & $IZ$ & $XZ$ & $IY$ & $YZ$ \\ \hline
$XX$ & --- & $iZX$ & $-iXY$ & $-iIY$ & $iXZ$ & $ZY$ \\
$YI$ &     & ---   & $YZ$   & $-iZZ$ & $YY$  & $IZ$ \\
$IZ$ &     &       & ---    & $XI$   & $-iIX$ & $YI$ \\
$XZ$ &     &       &        & ---    & $-iXX$ & $iZI$ \\
$IY$ &     &       &        &        & ---    & $iYX$ \\
$YZ$ &     &       &        &        &        & ---
\end{tabular}
\caption{Pairwise products of the six Pauli operators in
$\mathscr S=\{XX,YI,IZ,XZ,IY,YZ\}$, where $AB=A\otimes B$.
Each entry is the row operator multiplied by the column operator.
Up to an overall phase, the $\binom{6}{2}=15$ displayed products
exhaust all nonidentity two-qubit Pauli operators, without repetition.}
\label{tab:six-pauli-products}
\end{table}

From this observation, we can simplify the action of the channel on a Pauli matrix $A\neq I_4$. For Pauli operators $P,Q$, let $\Lambda_{P,Q}\in\{+1,-1\}$ denote their commutation sign, equal to $1$ if $P,Q$ commute and $-1$ otherwise. Let $Q,R\in\mathscr S$ be the unique Pauli matrices such that $A=QR$ up to a phase. Restricted to the ordered set $\mathscr S$, these signs form the commutation matrix $\Lambda$ given by:
\begin{eqnarray}
\Lambda = \begin{pmatrix}
1 & -1 & -1 & -1 & -1 & 1 \\
-1 & 1 & 1 & -1 & 1 & 1 \\
-1 & 1 & 1 & 1 & -1 & 1 \\
-1 & -1 & 1 & 1 & -1 & -1 \\
-1 & 1 & -1 & -1 & 1 & -1 \\
1 & 1 & 1 & -1 & -1 & 1
\end{pmatrix}.
\end{eqnarray}
With this notation, the action of the channel simplifies:
\begin{equation}
\begin{aligned}
\mathcal{F}(A) &= \frac{1}{6}\sum_{P\in\mathscr S}PQRP^\dagger \\
              &=\frac{A}{6}\sum_{P\in\mathscr S}\Lambda_{Q,P}\Lambda_{P,R}\\
              &=\frac{A}{6}(\Lambda^2)_{Q,R}\\
              &=\frac{\Lambda_{Q,R}}{3}A.
\end{aligned}
\label{eq:chaction}
\end{equation}
Since $Q\neq R$, the last equality follows from $\Lambda^2=4I_6+2\Lambda$, for which $(\Lambda^2)_{Q,R}=2\Lambda_{Q,R}$. Thus the sign of the eigenvalue of $\mathcal F$ on $A$ is precisely the commutation sign of the unique pair $Q,R$ generating $A$.

It is easy to see that the following is an isometric extension of the channel:
\begin{equation}
V_{\mathcal F} = \frac{1}{\sqrt6}\sum_{P\in\mathscr S} P\otimes\ket{P},
\end{equation}
with orthonormal states \(\{|P\rangle_F:P\in\mathscr S\}\). In consequence, the action of the complementary channel is given by 
\begin{equation}
\mathcal{F}^c(\rho)=\frac16\sum_{P,Q\in\mathscr S}\operatorname{tr}(QP\rho)\ketbra{P}{Q},
\end{equation}
with output dimension $6$. This channel encodes all Pauli expectations of the input state in the image of the channel.

The transpose-antidegrading map has a simple construction. Let \(|\Phi_4\rangle=\frac12\sum_{j=0}^{3}|j\rangle|j\rangle\), and map each environment basis state \(|P\rangle_F\) to \((I_4\otimes P)|\Phi_4\rangle\). These six vectors are orthonormal, so this map is an isometry. The degrading channel \(\mathcal D\) is given by tracing out its second four-dimensional system. Its action on a matrix unit is \(\mathcal D(|P\rangle\langle Q|)=(QP)^T/4\). 

The action of the degrading map composed with the complementary channel on a Pauli matrix $A\neq I_4$ is
\begin{equation}
    \begin{aligned}
        \mathcal{D}\circ\mathcal F^c(A)
        &= \mathcal{D}\left(\frac{1}{6}\sum_{P,S\in\mathscr S}\operatorname{tr}(ASP)\ketbra{P}{S}\right)\\
        &=\frac{1}{24}\sum_{P,S\in\mathscr S}\operatorname{tr}(ASP)(SP)^T\\
        &=\frac{\Lambda_{Q,R}}{3}A^T,
    \end{aligned}
\end{equation}
where $Q,R\in\mathscr S$ are again the unique pair generating $A$ up to phase. To see the last equality directly, write $QR=\alpha A$. Then $\alpha^2=\Lambda_{Q,R}$, because $(QR)^2=\Lambda_{Q,R}I_4$. In the sum only the two ordered pairs for which $SP$ is proportional to $A$, namely the orderings of $Q,R$, contribute. Their total contribution before the factor $1/24$ is $8\alpha^2 A^T=8\Lambda_{Q,R}A^T$. Thus, on every nonidentity two-qubit Pauli, $\mathcal D\circ\mathcal F^c$ has exactly the transpose of the action in \eqref{eq:chaction}. On the identity, both $\mathcal D\circ\mathcal F^c$ and $\mathsf T\circ\mathcal F$ map $I_4$ to $I_4$. Since the Pauli operators form a basis, $\mathcal D\circ\mathcal F^c=\mathsf T\circ\mathcal F$, and hence $\mathcal F$ is transpose antidegradable.

\section{An antidegradable Pauli channel}
\label{sec:anti}
Consider the following Pauli channel:
\begin{equation}
\begin{aligned}
\mathcal G(\rho)
={}&\frac14\rho
+\frac14(I\otimes Z)\rho(I\otimes Z)^\dagger\\
&+\frac1{16}
\sum_{P\in\{I,X,Y,Z\}}
\sum_{R\in\{X,Y\}}
(P\otimes R)\rho(P\otimes R)^\dagger.
\end{aligned}
\end{equation}

To understand the action of the channel, write
\begin{equation}
\rho=\sum_{t,s=0}^1\rho_{ts}^C\otimes\ketbra{t}{s}_D,
\end{equation}
where we denote the two input qubit registers by $C,D$ and the corresponding output registers by $C',D'$, and call them data and flag registers; thus $\mathcal G:CD\to C'D'$. For any operator $X^C$, we write $X^{C'}$ for its canonical copy on the isomorphic output register $C'$.

We can simplify the action of the channel as follows:
\begin{equation}
\frac14\left[
\rho+(I\otimes Z)\rho(I\otimes Z)^\dagger
\right]
=
\frac12\sum_{t=0}^1
\rho_{tt}^{C'}\otimes|t\rangle\langle t|_{D'}.
\end{equation}
and
\begin{equation}
\begin{aligned}
&\frac1{16}
\sum_{P\in\{I,X,Y,Z\}}
\sum_{R\in\{X,Y\}}
(P\otimes R)\rho(P\otimes R)^\dagger\\
&\qquad=
\frac18\sum_{t=0}^1
\sum_{P\in\{I,X,Y,Z\}}
P_{C'}\rho_{tt}^{C'} P_{C'}^\dagger
\otimes|\bar t\rangle\langle\bar t|_{D'}\\
&\qquad=
\frac12\sum_{t=0}^1
\operatorname{tr}(\rho_{tt}^C)\frac{I_{C'}}{2}
\otimes|\bar t\rangle\langle\bar t|_{D'}.
\end{aligned}
\end{equation}
where $\bar t=1-t$. Adding the two contributions gives
\begin{equation}
\mathcal G(\rho)
=
\frac12\sum_{t=0}^1
\left[
\rho_{tt}^{C'}\otimes|t\rangle\langle t|_{D'}
+
\operatorname{tr}(\rho_{tt}^C)\frac{I_{C'}}{2}
\otimes|\bar t\rangle\langle\bar t|_{D'}
\right].
\end{equation}
Intuitively, the channel first measures the input flag in the computational basis. Then, with uniform probability, it either transmits the data with the original flag, or replaces the data by a maximally mixed qubit and flip the flag.

We can now describe an isometric extension $V_{\mathcal G}:CD\to C'D'TJE$. Define
\begin{equation}
\begin{aligned}
V_{\mathcal G}|\psi\rangle_C|t\rangle_D
={}&\frac1{\sqrt2}
|\psi\rangle_{C'}|t\rangle_{D'}
|t\rangle_T|0\rangle_J|e\rangle_E\\
&+\frac12\sum_{j=0}^1
|j\rangle_{C'}|\bar t\rangle_{D'}
|t\rangle_T|j\rangle_J|\psi\rangle_E .
\end{aligned}
\end{equation}
where $|e\rangle_E$ is an erasure flag orthogonal to $|\psi\rangle_E$.

For the complementary channel, we restrict our attention to fixed inputs in the computational basis. For these inputs, it takes
the particularly simple form
\begin{equation}
\mathcal G^c(\tau^C\otimes|t\rangle\langle t|_D)
=
|t\rangle\langle t|_T\otimes
\left[
\frac12|0\rangle\langle0|_J\otimes|e\rangle\langle e|_E
+
\frac14 I_J\otimes\tau^E
\right].
\end{equation}
Here $\tau^E$ denotes the same qubit state as $\tau^C$, with the register relabelled from $C$ to $E$.

The environment can reproduce Bob's output as follows.
Measure $T$ in the computational basis to obtain $t$, and discard $J$.
If $E$ contains the erasure flag, prepare a maximally mixed data qubit on $C'$ and output flag $\bar t$ on $D'$. Otherwise, output the qubit in $E$ on $C'$ and flag $t$ on $D'$. Since $\mathcal G$ completely dephases the flag register $D$, its output depends only on the diagonal blocks $\rho_{tt}^C$. Measuring $T$ in the degrading map likewise removes coherences between different values of $t$, so the same procedure reproduces $\mathcal G(\rho)$ for an arbitrary input $\rho^{CD}$. Thus $\mathcal G$ is antidegradable.

\section{Superactivation of private capacity}
Let us first recall that transpose antidegradability of a channel
$\mathcal N$ is sufficient to rule out private communication.
For an input ensemble $\{p_u,\rho_u\}$, the classical label $U$ is
encoded in the possibly mixed state $\rho_u$ with probability $p_u$. Furthermore, we write
\begin{equation}
\rho_u^B=\mathcal N(\rho_u),
\qquad
\rho_u^E=\mathcal N^c(\rho_u),
\qquad
\overline\rho^B=\sum_u p_u\rho_u^B.
\end{equation}
The corresponding classical--quantum state at the receiver is
\begin{equation}
\rho^{UB}
=
\sum_u p_u|u\rangle\langle u|_U\otimes\rho_u^B.
\end{equation}
Its mutual information is
\begin{equation}
\mathbf I(U;B)
=
S(\overline\rho^B)-\sum_u p_uS(\rho_u^B),
\end{equation}
where $S(\rho)=-\operatorname{tr}(\rho\log_2\rho)$ is the
von Neumann entropy.
The environment's mutual information $\mathbf I(U;E)$ is defined
analogously using $\mathcal N^c$.
Maximizing the receiver's information advantage gives the
channel's private information,
\begin{equation}
P^{(1)}(\mathcal N)
=
\max_{\{p_u,\rho_u\}}
\left[\mathbf I(U;B)-\mathbf I(U;E)\right].
\end{equation}
Its private capacity is \cite{CaiWinterYeung2004,devetak2005private}:
\begin{equation}
P(\mathcal N)
=
\lim_{n\to\infty}
\frac1n P^{(1)}(\mathcal N^{\otimes n}).
\end{equation}

Transpose antidegradability means that there is a completely positive,
trace-preserving map $\mathcal D$ such that
\begin{equation}
\mathcal D\circ\mathcal N^c
=
\mathsf T_B\circ\mathcal N,
\end{equation}
where $\mathsf T_B$ denotes matrix transposition.
Transposition preserves the spectrum of each conditional output
state and of their average.
Data processing under $\mathcal D$ therefore gives
\begin{equation}
\mathbf I(U;B)\leq \mathbf I(U;E).
\end{equation}
The same argument applies to arbitrary joint input ensembles over
$n$ uses, since
\begin{equation}
\mathcal D^{\otimes n}\circ(\mathcal N^c)^{\otimes n}
=
\mathsf T_B^{\otimes n}\circ\mathcal N^{\otimes n}
\end{equation}
produces the full transpose of the receiver's joint output. In consequence the private capacity of $\mathcal N$ is zero.

This argument leaves an opening for joint use with another channel.
Transposition of the entire receiver output preserves its spectrum,
but transposition of only one part of a correlated quantum system
need not do so.
The entropy argument establishing zero private capacity therefore
does not by itself extend to an arbitrary quantum reference.

We now show that the transpose antidegradable $\mathcal F$ and the antidegradable helper
$\mathcal G$, each with zero private capacity, can jointly transmit
private classical information. We use $\mathcal G$ to highlight the superactivation effect with two Pauli channels. For the code below, the helper register $D$ is fixed to $\ketbra{0}$. On this subspace, the two branches of $\mathcal G$ coincide, up to output ancillas independent of the input letter, with those of a 50\% qubit erasure channel on $C$: with probability $1/2$ the input qubit is delivered to Bob, and with probability $1/2$ it is delivered to Eve. Consequently, the same mutual-information calculation also proves superactivation when $\mathcal G$ is replaced by the 50\% qubit erasure channel. We write $\mathcal F:KL\to K'L'$ for the main channel and $\mathcal F^c:KL\to F$ for its complementary channel.
The helper is $\mathcal G:CD\to C'D'$, with complementary channel $\mathcal G^c:CD\to TJE$.
Thus Bob receives $K',L',C',D'$, while Eve receives $F,T,J,E$.

Define the Bell states
\begin{equation}
|\phi^\pm\rangle
=
\frac{|00\rangle\pm|11\rangle}{\sqrt2},
\qquad
|\psi^\pm\rangle
=
\frac{|01\rangle\pm|10\rangle}{\sqrt2},
\end{equation}
and their density operators
$\Phi^\pm=|\phi^\pm\rangle\langle\phi^\pm|$ and
$\Psi^\pm=|\psi^\pm\rangle\langle\psi^\pm|$.
Alice chooses a uniformly distributed bit $U$ and prepares
\begin{equation}
\begin{aligned}
\sigma_0^{KLC}
&=
\frac12|0\rangle\langle0|_K\otimes\Phi^+_{LC}
+
\frac14|1\rangle\langle1|_K
\otimes(\Psi^-_{LC}+\Phi^-_{LC}),\\
\sigma_1^{KLC}
&=
\frac14|0\rangle\langle0|_K
\otimes(\Phi^+_{LC}+\Psi^+_{LC})
+
\frac12|1\rangle\langle1|_K\otimes\Phi^-_{LC}.
\end{aligned}
\end{equation}
She initializes the helper flag to $\ketbra{0}{0}$, so the actual input letters are
\begin{equation}
\widehat\sigma_u^{KLCD}
=
\sigma_u^{KLC}\otimes|0\rangle\langle0|_D,
\qquad u\in\{0,1\}.
\end{equation}

First apply $\mathcal F$, leaving $C$ unchanged, and write
\begin{equation}
\omega_u^{K'L'C}
=
(\mathcal F\otimes\operatorname{id}_C)(\sigma_u),
\qquad
\omega_u^{FC}
=
(\mathcal F^c\otimes\operatorname{id}_C)(\sigma_u).
\end{equation}
Both input letters have $KL$ marginal $I_{KL}/4$.
Since $\mathcal F$ is unital and its six equally weighted Pauli
Kraus operators are orthogonal under the Hilbert--Schmidt inner product,
\begin{equation}
\omega_0^{K'L'}=\omega_1^{K'L'}=\frac{I_{K'L'}}4,
\qquad
\omega_0^F=\omega_1^F=\frac{I_F}{6}.
\end{equation}
Thus neither $K'L'$ nor $F$ alone contains information about $U$.
The information is contained in their correlations with $C$.

We now apply the helper to $C,D$.
With probability $1/2$, Bob receives the input qubit $C$ on the output register $C'$ and Eve receives the erasure flag.
With probability $1/2$, Eve receives $C$ on $E$ and Bob receives a maximally mixed qubit on $C'$.
We write $\omega_u^{K'L'C'}$ for $\omega_u^{K'L'C}$ with $C$ relabelled as $C'$, and $\omega_u^{FE}$ for $\omega_u^{FC}$ with $C$ relabelled as $E$.
The resulting state is therefore
\begin{equation}
\beta_u^{K'L'C'D'}
=
\frac12\omega_u^{K'L'C'}\otimes|0\rangle\langle0|_{D'}
+
\frac12\omega_u^{K'L'}\otimes\frac{I_{C'}}{2}
\otimes|1\rangle\langle1|_{D'}
\end{equation}

Similarly,
\begin{equation}
\begin{aligned}
\xi_u^{TFEJ}
=
|0\rangle\langle0|_T\otimes
\biggl[
&\frac12\omega_u^F
\otimes|e\rangle\langle e|_E
\otimes|0\rangle\langle0|_J+\frac12\omega_u^{FE}\otimes\frac{I_J}{2}
\biggr],
\end{aligned}
\end{equation}
where $T,F,E,J$ registers have been reordered for clarity.

For each recipient, the two terms occupy orthogonal subspaces and
have probabilities $1/2$ independent of $U$.
Consequently,
\begin{equation}
\begin{aligned}
\mathbf I(U;K'L'C'D')
&=
\frac12 \mathbf I(U;K'L'C')+\frac12 \mathbf I(U;K'L'),\\
\mathbf I(U;TFEJ)
&=
\frac12 \mathbf I(U;F)+\frac12 \mathbf I(U;FE).
\end{aligned}
\end{equation}
The fixed states of $T$ and $J$, and the independent maximally mixed
states, do not contribute to these quantities.
Since $\mathbf I(U;K'L')=\mathbf I(U;F)=0$ and mutual information is unchanged by the relabelling $C\to C'$ and $C\to E$, the private information for the input ensemble simplifies to
\begin{equation}
\mathbf I(U;K'L'C'D')-\mathbf I(U;TFEJ)=\frac12\left[\mathbf I(U;K'L'C)-\mathbf I(U;FC)\right].
\end{equation}

Each Pauli error of $\mathcal F$ maps a computational-basis state on $K$ and a Bell state on $LC$ to another vector of the same product basis, up to a phase. Since both input letters are diagonal in this basis, the states $\omega_0^{K'L'C}$ and $\omega_1^{K'L'C}$ remain diagonal in the basis consisting of a computational-basis state on $K'$ and a Bell state on $L'C$. Their diagonal coefficients are listed in Table~\ref{tab:bob-output}.

\begin{table}[t]
\centering
\renewcommand{\arraystretch}{1.3}
\begin{tabular}{c|cc}
Basis ket
&
$(\mathcal F\otimes\operatorname{id}_C)(\sigma_0)$
&
$(\mathcal F\otimes\operatorname{id}_C)(\sigma_1)$
\\ \hline
$\lvert 0,\phi^+\rangle$ & $\frac{1}{12}$ & $\frac{1}{6}$ \\
$\lvert 1,\phi^-\rangle$ & $\frac{1}{6}$ & $\frac{1}{12}$ \\
$\lvert 0,\phi^-\rangle$ & $\frac{1}{6}$ & $\frac{1}{6}$ \\
$\lvert 1,\phi^+\rangle$ & $\frac{1}{6}$ & $\frac{1}{6}$ \\
$\lvert 0,\psi^+\rangle$ & $\frac{1}{12}$ & $0$ \\
$\lvert 1,\psi^-\rangle$ & $0$ & $\frac{1}{12}$ \\
$\lvert 0,\psi^-\rangle$ & $\frac{1}{6}$ & $\frac{1}{6}$ \\
$\lvert 1,\psi^+\rangle$ & $\frac{1}{6}$ & $\frac{1}{6}$
\end{tabular}
\caption{Diagonal coefficients of
$(\mathcal F\otimes\operatorname{id}_C)(\sigma_0)$ and
$(\mathcal F\otimes\operatorname{id}_C)(\sigma_1)$
in the basis
$\{\lvert k\rangle_{K'}\lvert\beta\rangle_{L'C}\}$,
where $k\in\{0,1\}$ and
$\beta\in\{\phi^+,\phi^-,\psi^+,\psi^-\}$.}
\label{tab:bob-output}
\end{table}

Reading the probabilities in Table~\ref{tab:bob-output} gives
\begin{equation}
\label{eq:ibob}
\mathbf I(U;K'L'C)=\frac14\log_2\frac43.
\end{equation}

To evaluate Eve's mutual information, we first simplify her states
by a unitary on $FC$.
Write each Pauli error of $\mathcal F$ as $A_{K\to K'}\otimes Q_{L\to L'}$, where the subscripts indicate the corresponding input and output registers, and let $|AQ\rangle_F$ denote its environment label.
All sums over $A,Q$ below run over the six errors of $\mathcal F$.
The Pauli dilation is
\begin{equation}
V_{\mathcal F}
=
\frac1{\sqrt6}
\sum_{A,Q}
(A_{K\to K'}\otimes Q_{L\to L'})\otimes|AQ\rangle_F.
\end{equation}
Consider the controlled unitary
\begin{equation}
W_{FC}
=
\sum_{A,Q}
|AQ\rangle\langle AQ|_F\otimes Q_C^*,
\end{equation}
where the star denotes complex conjugation in the computational basis.

Every Bell state can be written, up to a global phase, as
\begin{equation}
\begin{aligned}
|\beta_R\rangle_{LC}&=(I_L\otimes R_C)|\phi^+\rangle_{LC},\\
|\beta_R\rangle_{L'C}&=(I_{L'}\otimes R_C)|\phi^+\rangle_{L'C},
\end{aligned}
\qquad R\in\{I,X,Y,Z\}.
\end{equation}
For every Pauli $Q$, the maximally entangled state satisfies $(Q_{L'}\otimes Q_C^*)|\phi^+\rangle_{L'C}=|\phi^+\rangle_{L'C}$. Together with $QR=\Lambda_{Q,R}RQ$, this gives
\begin{equation}
(Q_{L'}\otimes Q_C^*)|\beta_R\rangle_{L'C}
=
\Lambda_{Q,R}|\beta_R\rangle_{L'C}.
\end{equation}
Thus, after the dilation and the controlled unitary, an input component
$|k\rangle_K|\beta_R\rangle_{LC}$ has the factorized form
\begin{equation}
|\gamma_{k,R}\rangle_{K'F}\otimes|\beta_R\rangle_{L'C},
\qquad
|\gamma_{k,R}\rangle_{K'F}
=
\frac1{\sqrt6}\sum_{A,Q}
\Lambda_{Q,R}\bigl(A_{K\to K'}|k\rangle_K\bigr)\otimes|AQ\rangle_F,
\end{equation}
with the tensor factors grouped as $K'F$ and $L'C$.
Tracing out $K',L'$ therefore leaves
$\operatorname{tr}_{K'}(|\gamma_{k,R}\rangle\langle\gamma_{k,R}|)
\otimes I_C/2$.

The same unitary works for every component of both input letters.
Consequently, there are states $\eta_u^F$ such that
\begin{equation}
W_{FC}\omega_u^{FC}W_{FC}^\dagger
=
\eta_u^F\otimes\frac{I_C}{2},
\qquad u\in\{0,1\}.
\end{equation}
Writing $\overline\eta=(\eta_0+\eta_1)/2$, unitary invariance and
additivity of entropy give
\begin{equation}
\mathbf I(U;FC)
=
S(\overline\eta)
-\frac12S(\eta_0)-\frac12S(\eta_1).
\end{equation}
The joint unitary has transferred the information in the correlations
between $F$ and $C$ into $F$, leaving $C$ in a fixed maximally mixed state.

We now compute the states $\eta_u$.
For brevity, write $[v]=|v\rangle\langle v|$.
Introduce the following orthonormal vectors in $F$:
\begin{equation}
\begin{aligned}
|a_\pm\rangle
&=
\frac{|IZ\rangle\pm|IY\rangle}{\sqrt2},\\
|s\rangle
&=
\frac{|XX\rangle+i|YI\rangle}{\sqrt2},\\
|p\rangle
&=
\frac{|XZ\rangle+i|YZ\rangle}{\sqrt2},
\qquad
|m\rangle
=
\frac{|XZ\rangle-i|YZ\rangle}{\sqrt2}.
\end{aligned}
\end{equation}
The errors $IZ,IY$ preserve the computational-basis value of the input register $K$ at the output $K'$, while the other four errors flip it.
These two alternatives have orthogonal outputs on $K'$, so their cross terms disappear when $K'$ is traced out.
Their probabilities are respectively $2/6=1/3$ and $4/6=2/3$.

Applying the expression for $|\gamma_{k,R}\rangle$ gives
Table~\ref{tab:eve-components}.
For example, the component $|0,\phi^+\rangle$ gives
\begin{equation}
|\gamma_{0,I}\rangle_{K'F}
=
\sqrt{\frac13}|0\rangle_{K'}|a_+\rangle_F
+
\sqrt{\frac23}|1\rangle_{K'}
\frac{|s\rangle_F+|p\rangle_F}{\sqrt2}.
\end{equation}

\begin{table}[t]
\centering
\begin{tabular}{c|c}
Input component on $KLC$ & Corrected state on $F$
\\ \hline
$|0,\phi^+\rangle$
&
$\displaystyle
\frac13[a_+]
+\frac23\left[\frac{s+p}{\sqrt2}\right]$
\\[1.2ex]
$|0,\psi^+\rangle$
&
$\displaystyle
\frac13[a_+]
+\frac23\left[\frac{s-p}{\sqrt2}\right]$
\\[1.2ex]
$|1,\psi^-\rangle$
&
$\displaystyle
\frac13[a_-]
+\frac23\left[\frac{s+m}{\sqrt2}\right]$
\\[1.2ex]
$|1,\phi^-\rangle$
&
$\displaystyle
\frac13[a_-]
+\frac23\left[\frac{s-m}{\sqrt2}\right]$
\end{tabular}
\caption{Environment states after applying $W_{FC}$ and removing
the common factor $I_C/2$. Every ket appearing inside a projector
is normalized.}
\label{tab:eve-components}
\end{table}

Combining these states with the probabilities defining
$\sigma_0,\sigma_1$ gives
\begin{equation}
\begin{aligned}
\eta_0
&=
\frac16\bigl([a_+]+[a_-]\bigr)
+\frac13\left[\frac{s+p}{\sqrt2}\right]
+\frac16[s]+\frac16[m],\\
\eta_1
&=
\frac16\bigl([a_+]+[a_-]\bigr)
+\frac13\left[\frac{s-m}{\sqrt2}\right]
+\frac16[s]+\frac16[p].
\end{aligned}
\end{equation}
These states are unitarily equivalent: fix $s,a_+,a_-$ and send
$p\mapsto-m$ and $m\mapsto p$.
Hence $S(\eta_0)=S(\eta_1)$, and
\begin{equation}
\mathbf I(U;FC)=S(\overline\eta)-S(\eta_0).
\end{equation}

The contribution on $\operatorname{span}\{a_+,a_-\}$ is identical
in both states, with eigenvalues $1/6,1/6$.
For $\eta_0$, the vector $m$ also has eigenvalue $1/6$.
The remaining block, in the basis $(s,p)$, is
\begin{equation}
M_0
=
\frac1{12}
\begin{pmatrix}
4&2\\
2&2
\end{pmatrix}.
\end{equation}

For the average state, the vector $(p+m)/\sqrt2$ has eigenvalue $1/6$.
In the orthogonal basis
$\bigl(s,(p-m)/\sqrt2\bigr)$, the remaining block is
\begin{equation}
\overline M
=
\frac1{12}
\begin{pmatrix}
4&\sqrt2\\
\sqrt2&2
\end{pmatrix}.
\end{equation}
There is also one common zero eigenvalue, corresponding to the
unused direction in $F$.
Thus only the eigenvalues of these two $2\times2$ blocks contribute
to the entropy difference:
\begin{equation}
\operatorname{spec}(M_0)
=
\left\{\frac{3-\sqrt5}{12},\frac{3+\sqrt5}{12}\right\},
\qquad
\operatorname{spec}(\overline M)
=
\left\{\frac{3-\sqrt3}{12},\frac{3+\sqrt3}{12}\right\}.
\end{equation}
Each pair has total weight $1/2$, so
\begin{equation}
\mathbf I(U;FC)
=
\frac12\left[
h_2\left(\frac{3-\sqrt3}{6}\right)
-
h_2\left(\frac{3-\sqrt5}{6}\right)
\right],
\end{equation}
where \(h_2(x)=-x\log_2x-(1-x)\log_2(1-x)\). 

Finally, the helper delivers the input qubit $C$ to Eve on register $E$ with probability $1/2$, while the other branch carries no information about $U$.
Therefore her mutual information for the full product channel is
\begin{equation}
\mathbf I(U;TFEJ)
=
\frac12 \mathbf I(U;FE)
=
\frac12 \mathbf I(U;FC)
=
\frac14\left[
h_2\left(\frac{3-\sqrt3}{6}\right)
-
h_2\left(\frac{3-\sqrt5}{6}\right)
\right].
\label{eq:ieve}
\end{equation}

Plugging \eqref{eq:ibob} and \eqref{eq:ieve} we obtain
\begin{equation}
\begin{aligned}
P(\mathcal F\otimes\mathcal G)
&\geq P^{(1)}(\mathcal F\otimes\mathcal G)\\
&\geq \frac18\log_2\frac43
-\frac14\left[
 h_2\left(\frac{3-\sqrt3}{6}\right)
-h_2\left(\frac{3-\sqrt5}{6}\right)
\right]\\
&\approx 0.00338974>0.
\end{aligned}
\label{eq:private-advantage}
\end{equation}
Together with $P(\mathcal F)=P(\mathcal G)=0$, this establishes private-capacity superactivation for two two-qubit Pauli channels. It also certifies that $\mathcal F$ is not antidegradable: otherwise $\mathcal F\otimes\mathcal G$ would be antidegradable and hence have zero private capacity. More generally, $\mathcal F$ is not anti cLN: since $\mathcal G$ is antidegradable and therefore anti cLN, tensor-product stability would otherwise make $\mathcal F\otimes\mathcal G$ anti cLN as well, again forcing zero private capacity.

\section{Discussion}
In summary, we have shown that $P(\mathcal F)=P(\mathcal G)=0$ but $P(\mathcal F\otimes\mathcal G)>0$. Our construction realizes superactivation using two two-qubit Pauli channels and a simple binary code. The Pauli structure also provides a direct proof and a simple interpretation of transpose antidegradability. The six errors defining \(\mathcal F\) have pairwise products that exhaust all nonidentity two-qubit Paulis. Consequently, the complementary, transpose-degradable channel \(\mathcal F^c\) encodes every input Pauli expectation in a distinct pair of off-diagonal matrix elements, while \(\mathcal F\) simply rescales each nonidentity Pauli operator by \(\pm 1/3\). The transpose-antidegrading map embeds the six environment labels into orthogonal maximally entangled states and traces out one subsystem. The proof of transpose antidegradability follows directly from the commutation relations of the Pauli matrices.

Let $\mathcal E_p^{(d)}$ denote the $d$-dimensional erasure channel with erasure probability $p$. The same private code also gives a finite-dimensional quantum-capacity superactivation through the Smith--Yard coherentization construction \cite{Smithyard}. Each of the two letters $\widehat\sigma_0$ and $\widehat\sigma_1$ has rank three. Choose purifications $|\widehat\sigma_u\rangle^{KLCDR}$ whose purifying systems occupy orthogonal three-dimensional subspaces of a six-dimensional register $R$, and define the mixed input
\begin{equation}
\rho^{KLCDR}
=\frac12\sum_{u=0}^1
|\widehat\sigma_u\rangle\langle\widehat\sigma_u|^{KLCDR}.
\end{equation}
Equivalently, a purification of this input is
\begin{equation}
|\Omega\rangle^{UKLCDR}
=\frac{1}{\sqrt2}\sum_{u=0}^1
|u\rangle_U|\widehat\sigma_u\rangle^{KLCDR},
\end{equation}
which is the coherent version of the classical ensemble used above. Sending $R$ through $\mathcal E_{1/2}^{(6)}$, the Smith--Yard identity gives
\begin{equation}
\begin{aligned}
Q(\mathcal F\otimes\mathcal G\otimes\mathcal E_{1/2}^{(6)})
&\geq \mathbf I_c\!\left(\rho,\mathcal F\otimes\mathcal G\otimes\mathcal E_{1/2}^{(6)}\right)\\
&=\frac12\left[\mathbf I(U;K'L'C'D')-\mathbf I(U;TFEJ)\right]\\
&>0,
\end{aligned}
\end{equation}
where $\mathbf I_c$ denotes coherent information and the last line is positive by \eqref{eq:private-advantage}.

Define $\mathcal H=\mathcal G\otimes\mathcal E_{1/2}^{(6)}$. The 50\% erasure channel is self-complementary and hence antidegradable; tensor products of antidegradable channels are antidegradable. Therefore $\mathcal H$ is antidegradable and $P(\mathcal H)=Q(\mathcal H)=0$. Since $Q\leq P$, $P(\mathcal F)=0$ also implies $Q(\mathcal F)=0$. Thus the pair $\mathcal F,\mathcal H$ exhibits a stronger form of quantum-capacity superactivation where both channels have zero \textit{private} capacity. 

While finalizing the preparation of the manuscript we became aware of an analogous independent result by Zhu and Wang \cite{zhu2026private}. Both constructions are similar; they consist of an antidegradable channel and a transpose-antidegradable channel. However, our transpose antidegradable channel has input, output and environment dimensions (4,4,6) while the one in Zhu and Wang is (4,4,8). The possibility of transpose antidegradability in smaller dimensions remains open as the minimum possible triples are $
 \{(2,4,5),\ (3,3,4),\ (4,2,5)\}$ \cite{qiqcop}.

Although the separation between transpose degradability and degradability remained elusive for many years, the examples now available suggest that it may be more widespread than this history would suggest. This motivates a systematic investigation of the structure of transpose-degradable channels that are not degradable, and the identification of general families beyond the present constructions. A related operational question is whether every transpose-antidegradable channel that is not antidegradable can superactivate private capacity with a suitable antidegradable channel. Equivalently, does every such channel \(\mathcal N\) admit an antidegradable partner \(\mathcal A\) for which \(P(\mathcal N\otimes\mathcal A)>0\)? Both our construction and that of \cite{zhu2026private} exhibit this behavior, raising the possibility that superactivation is a general consequence of the separation between transpose antidegradability and antidegradability.
\section*{Acknowledgements}
The authors used OpenAI's ChatGPT (GPT-5.6 Sol Pro) in the development of this work. The system was used to explore candidate constructions and assist with drafting the manuscript. The authors assume full responsibility for every claim, calculation, and conclusion.

The authors were supported by Japan’s Council for Science, Technology and Innovation (CSTI) under the Cross-ministerial Strategic Innovation Promotion Program (SIP) for ``Promoting the application of advanced quantum technology platforms to social issues" (Grant JPJ012367).

\appendix

\bibliographystyle{IEEEtran}
\bibliography{reference}

\end{document}